\documentclass{ragtime} % NO optional arguments are required
\title[The gyraton solutions on generalized Melvin universe with cosmological constant]%
      {The gyraton solutions on generalized\\ Melvin universe with cosmological constant}

\author[H.Kadlecov\v{a} % Run. head authors: separate names with commas,
        and P. Krtou\v{s}]%    Now let's start the paper title authors:
       {Hedvika Kadlecov\'{a}\at[]{a} % Makes referencing superscript `1'
        and Pavel Krtou\v{s}\at{b}\\% Termination of authors' block; if
        Institute of Theoretical Physics, Faculty of Mathematics and Physics,
        Charles University,\splitins[1]% This is how to break an
        V Hole\v{s}ovi\v{c}k\'{a}ch~2, 180\,00 Prague,
        Czech Republic.\\% Termination of the first affiliation.
        \ins{a}\Email{hedvika.kadlecova@centrum.cz},
        \ins{b}\Email{Pavel.Krtous@utf.mff.cuni.cz}} % This is how to present E-mail.

\coentry{Z. Stuchl\'{\i}k, J. Kov\'a\v{r} and J. Cimrman}%
       {Equilibrium of spinning test particles in equatorial plane\par
        of Kerr--de~Sitter spacetimes}

\newcommand{\rot}{\mathrm{rot}\,}
\newcommand{\laplace}{{\vartriangle}\,}

\newcommand{\epso}{\varepsilon_{\mathrm{o}}}

\newcommand{\EM}{{\scriptscriptstyle\mathrm{EM}}}
\newcommand{\gyr}{{\scriptscriptstyle\mathrm{gyr}}}
\newcommand{\vrho}{\varrho_{\scriptscriptstyle\mathrm{EM}}}
\newcommand{\vs}{{\bf s}}

\DeclareMathAccent{\frmarr}{\mathord}{letters}{"7E}

\newcommand{\trgrad}[1][]{\mathrm{d}_{\mspace{1mu}#1}\mspace{-1mu}}

\DeclareMathOperator{\trdiv}{div}

\begin{document}

% Citation of references in abstract should generally be avoided to
% ensure self-consistency of the abstract.  If you do insist on citation(s)
% within the abstract, you should use the \bibentry command, which forces
% the _complete_bibliographic_entry_ to appear in the abstract.
% With the `nonatbib' optional class argument this feature is not available.
\begin{abstract}
 We present and analyze new exact gyraton solutions of algebraic type II on generalized
 Melvin universe of type D which admit non--vanishing cosmological constant $\Lambda$. We show that it generalizes both, gyraton solutions on Melvin and on direct product spacetimes.
 When we set $\Lambda=0$ we get solutions on Melvin spacetime and for $\Sigma=1$ we obtain solutions on direct product spacetimes. We demonstrate that the solutions are member of the Kundt family of spacetimes as its subcases. We show that the Einstein equations reduce to a set of equations on the transverse 2-space. We also discuss the polynomial scalar invariants which are non--constant in general but constant
 for sub--solutions on direct product spacetimes.
\end{abstract}

% e.g., Levi-Civita, which is a single person):
\begin{keywords}
Gyraton solutions~-- Melvin universe~-- cosmological constant~--
Kundt family~-- direct product spacetimes~-- constant polynomial scalar invariants~-- Einstein equations
\end{keywords}

% It is good to provide as many as \label's possible, but never start the
% key with a numeral.  This makes problems with pdflatex processing.
\section{Introduction}\label{intro}%%%%%%%%%%%%%%%%%%%%%%%%%%%%%%%%%%%%%%%%
In \cite{Kadlecova:2009:PHYSR4:} and \cite{Kadlecova:2010:PHYSR4:} we have investigated the gyraton solutions on direct product spacetimes and gyraton solutions on Melvin universe. These solutions are of algebraic type II. In this work we present the gyraton solutions on Melvin universe with the cosmological constant. 

We present our ansatz for the gyraton metric on generalized Melvin universe and the generalized electromagnetic tensor. We briefly review the derivation of the Einstein--Maxwell equations. The source--free Einstein equations determine the functions $\Sigma$ and $S$, in particular, there exists a relation between them. Next we derive the non--trivial source equations. The Einstein--Maxwell equations do decouple for the gyraton metric on generalized Melvin universe as for its subcase solutions on Melvin and on direct product spacetimes. Next, we focus on interpretation of our solutions. Especially, we discuss the geometry of the transverse metric of the generalized Melvin universe in detail for different values of the cosmological constant. We show explicitly that the Melvin universe and direct product spacetimes are special cases of our solutions. We also discuss the properties of the scalar polynomial invariants which are functions of $\rho$ but for subcase solutions on direct product spacetimes $(\Sigma=1)$ the invariants are constant.
%%%%%%%%%%%%%%%%%%%%%%%%%%%%%%%%%%%%%%%%%%%%%%%%%%%%%%%%%%%%%%%%%%%%%%%%%%%%%%%%%%%%%%%%%
%%%%%%%%%%%%%%%%%%%%%%%%%%%%%%%%%%%%%%%%%%%%%%%%%%%%%%%%%%%%%%%%%%%%%%%%%%%%%%%%%%%%%%%%%
\section{The ansatz for the gyratons on generalized Melvin universe}\label{sc:def}

The ansatz for the gyraton metric on the generalized Melvin spacetime is the following,
\begin{equation}\label{s1w}
{\bf g}=-2\Sigma^2 H \trgrad u^2-\Sigma^2\trgrad u\vee\trgrad v + {\bf q} +\Sigma^2 \trgrad u\vee {\bf a},
\end{equation}
where we have introduced the 2--dimensional transversal metric ${\bf q}$  on transverse spaces $u,\,v=$constant as
\begin{equation}\label{trmetric4}
{\bf q}=\Sigma^2\trgrad\rho^2+\frac{S(\rho)^2}{\Sigma^2}\trgrad \phi^2.
\end{equation}
We have assumed that the metric \eqref{s1w} belongs to the Kundt class of spacetimes and
that the transversal metric ${\bf q}$ has one Killing vector $\mathcal{L}_{\frac{\partial}{\partial \phi}}{\bf q}=0.$
The metric \eqref{s1w} represents gyraton propagating on the background which is formed by generalized Melvin
spacetime. The metric \eqref{s1w} generalizes only the transversal metric therefore the algebraical type is $II$ as
for the gyraton on the Melvin spacetime \cite{Kadlecova:2010:PHYSR4:}, the NP quantities are listed in \cite{Kadlecova:2013:}.

We have generalized the transversal metric for the Melvin universe by assuming
general function $S=S(\rho)$ instead of the simple coordinate $\rho$ in front of the term $\trgrad\phi^2$, see \cite{Kadlecova:2010:PHYSR4:}.
We will show that these general functions $\Sigma(\rho)$ and $S(\rho)$ are determined by the Einstein--Maxwell equations and have proper interpretation. The presence of cosmological constant $\Lambda$ is not allowed for the solution on pure Melvin background \cite{Kadlecova:2010:PHYSR4:}.

The transverse space is covered by two  spatial coordinates ${x^i}$ $(i=\rho,\,\phi)$ and it is convenient to introduce suitable notation on it, technical details can be found in \cite{Kadlecova:2013:}.
The function $H(u,v,{\bf x})$ in the metric \eqref{s1w} can depend on all coordinates, but the functions ${a(u,{\bf x})}$ are ${\mbox{${v}$-independent}}$.

The derivation of the Einstein--Maxwell equations is almost identical with the previous paper \cite{Kadlecova:2010:PHYSR4:} therefore we will describe the derivation of Einstein--Maxwell equations very briefly.

The metric should satisfy the Einstein equations with cosmological constant $\Lambda$ and with a  stress-energy tensor generated by the electromagnetic field of the background Melvin spacetime  ${\bf T}^\EM$ and the gyratonic source ${\bf T}^{\gyr}$ as\footnote{$\varkappa=8\pi G$ and $\epso$ are gravitational and electromagnetic constants. There are two general choices of geometrical units: the gaussian with $\varkappa=8\pi$ and $\varepsilon_{\rm o}=1/4\pi$, and SI-like with $\varkappa=\varepsilon_{\rm o}=1$.}

\begin{equation}\label{EinsteinEqqw}
{\bf G}+\Lambda\,{\bf g}=\varkappa \bigl( {\bf T}^\EM+{\bf T}^{\gyr}\bigr)\;.
\end{equation}
We assume the electromagnetic field is given by
\begin{equation}\label{realFF}
{\bf F}=E\trgrad v \wedge\trgrad u+\frac{B}{\Sigma^2}{\boldsymbol{\epsilon}}+\trgrad u\wedge (E\,\vs-B{*(\vs-{\bf a})})\;,
\end{equation}
where  $E$ and $B$ are parameters of electromagnetic field.
The self--dual complex form of the Maxwell\footnote{We will follow the notation of \cite{Step:2003:Cam:}. Namely, $\boldsymbol{\mathcal{F}}\equiv {\bf F}+i{{\star}\bf{F}}$ is complex self--dual Maxwell tensor, where the 4--dimesional Hodge dual is ${\star}{F}_{\mu\nu}=\frac{1}{2}\varepsilon_{\mu\nu\rho\sigma}{F}^{\rho\sigma}$. The self--dual condition reads ${{\star}\boldsymbol{\mathcal{F}}}=-i\boldsymbol{\mathcal{F}}$. The orientation of the 4--dimensional Levi--Civita tensor
is fixed by the sign of the component $\varepsilon_{vu\rho\phi}=S\Sigma^2$.
The energy--momentum tensor of the electromagnetic field is given by $T_{\mu\nu}=\frac{\varepsilon_{\rm o}}{2}\mathcal{F}_{\mu}^{\rho}\overline{\mathcal{F}}_{\nu\rho}$.}
 tensor is
\begin{equation}\begin{split}\label{EMFa}
\boldsymbol{\mathcal{F}}&=\mathcal{B}(\trgrad v\wedge\trgrad u - \frac{i}{\Sigma^2}{\boldsymbol \epsilon}+\trgrad u\wedge[\vs+i{*(\vs}-{\boldsymbol{a}})]),
\end{split}\end{equation}
for details see \cite{Kadlecova:2010:PHYSR4:}.

We have denoted the complex constant $\mathcal{B}=E+iB,$
and we have introduced a constant $\vrho$,
\begin{equation}\label{rhodeff}
\vrho=\frac{\varkappa\epso}{2}(E^2+B^2).
\end{equation}

We define the gyratonic matter only on a phenomenological level as
\begin{equation}\label{mm7}
\varkappa\, {\bf T}^{\gyr}=j_{u}\,\trgrad u^2+\trgrad u\vee {\bf j}\;,
\end{equation}
where the source functions ${j_u(v,u,{\bf x})}$ and ${j(v,u,{\bf x})}$.
We assume that the gyraton stress-energy tensor is locally conserved,
\begin{equation}\label{gyrcon}
  \nabla {\cdot} {\bf T}^{\gyr}=0\;.
\end{equation}

To conclude, the fields are characterized by functions ${\Sigma}$, $S$, ${H}$, ${\bf a}$, and ${\vs}$ which must be determined by the field equations and the gyraton sources ${j_u}$ and ${\bf j}$ and the constants ${E}$ and ${B}$ of the background electromagnetic field are prescribed.
%%%%%%%%%%%%%%%%%%%%%%%%%%%%%%%%%%%%%%%%%%%%%%%%%%%%%%%%%%%%%%%%%
\section{The Einstein--Maxwell field equations}\label{scc:fequationsss}
First, we will start to solve the Maxwell equations, it is sufficient to calculate the cyclic Maxwell equation for the self--dual Maxwell tensor \eqref{EMFa}
\begin{equation}\begin{aligned}\label{MXECC}
  0=\trgrad {\mathcal{F}} =\mathcal{B}&\left\{\partial_v(\vs+i{*(\vs-{\bf a})})\, \trgrad v\wedge\trgrad u\wedge\trgrad {\bf x}-[\rot\vs+i\,\trdiv(\vs-{\bf a})]\, \trgrad u \wedge \boldsymbol{\epsilon}\right\}\;.
\end{aligned}\end{equation}

From the real part we immediately get that the 1-forms ${\vs}$ is ${v}$-independent, and rotation free $ \rot\vs=0\;$.
From imaginary part it follows that the 1--form ${\bf a}$ is also independent and it satisfies $ \trdiv(\vs-{\bf a}) = 0\;. $
%The equations \eqref{pot1} and \eqref{pot2} determine the potentials which we will be discussed in Section \ref{sources}.

\subsection{The trivial  Einstein--Maxwell equations--determining the function $\Sigma$ and $S$}
Next we will derive the Einstein--Maxwell equations from the Einstein tensor and the electromagnetic stress-energy tensor, which are listed in \cite{Kadlecova:2013:}.

First we will  solve the equations which are source free and we will be able to determine
the analytic formula for the functions $\Sigma$ and $S$.

The first equation we obtain from the $vu$-component,
\begin{equation}\label{r1}
-\frac{(\Sigma_{,\rho})^2}{\Sigma^2}+2\frac{\Sigma_{,\rho}}{\Sigma}\frac{S_{,\rho}}{S}-\frac{S_{,\rho\rho}}{S}=\Lambda\Sigma^2+\frac{\vrho}{\Sigma^2},
\end{equation}
the next two equations we get from the transverse diagonal components $\rho\rho$ and $\phi\phi$,
\begin{align}
-\frac{(\Sigma_{,\rho})^2}{\Sigma^2}+2\frac{\Sigma_{,\rho}}{\Sigma}\frac{S_{,\rho}}{S}+\partial^2_{v}H&=-\Lambda\Sigma^2+\frac{\vrho}{\Sigma^2}\;,\label{r2}\\
-\frac{(\Sigma_{,\rho})^2}{\Sigma^2}+2\frac{\Sigma_{,\rho\rho}}{\Sigma}+\partial^2_{v}H&=-\Lambda\Sigma^2+\frac{\vrho}{\Sigma^2}\;.\label{r3}
\end{align}
When we compare the equation \eqref{r2} and \eqref{r3} we immediately get
the relation between the functions $\Sigma$ and $S$, as $
\Sigma_{,\rho}\frac{S_{,\rho}}{S}=\Sigma_{,\rho\rho},$
and thus we are able to determine their explicit relation ($\Sigma_{,\rho}\neq0$) as
\begin{equation}\label{s3q}
\Sigma_{,\rho}=\gamma S,
\end{equation}
where $\gamma$ is an integration constant.

After substituting the relation \eqref{s3q} into equation \eqref{r1} then we get equation
\begin{equation}
-\frac{(\Sigma_{,\rho})^2}{\Sigma^2}+2\frac{\Sigma_{,\rho\rho}}{\Sigma}+\frac{\Sigma_{,\rho\rho\rho}}{\Sigma_{,\rho}}=\Lambda\Sigma^2+\frac{\vrho}{\Sigma^2}\;,\label{r4}
\end{equation}
which will be useful later.

To determine the function $H$ it is useful to substitute \eqref{s3q} into the equation \eqref{r3} and then multiply it by $\frac12\frac{\Sigma}{\Sigma_{,\rho}}$,
we get
\begin{equation}\label{r5}
\frac12(\partial^2_{v}H)_{,\rho}\frac{\Sigma}{\Sigma_{,\rho}}-2\frac{\Sigma_{,\rho\rho}}{\Sigma}+\frac{(\Sigma_{,\rho})^2}{\Sigma^2}+\frac{\Sigma_{,\rho\rho\rho}}{\Sigma_{,\rho}}=-\Lambda\Sigma^2-\frac{\vrho}{\Sigma^2}\;.
\end{equation}
Now, we add the equation \eqref{r1} to \eqref{r5} and  obtain, $
\frac12(\partial^2_{v}H)_{,\rho}\frac{\Sigma}{\Sigma_{,\rho}}=0, $
then for $\Sigma_{,\rho}\neq 0$ we can write that $ \partial^2_{v} H  =-\alpha\;, $
where $\alpha$ is a constant.

Thus the metric function ${H}$ has a structure
\begin{equation}\label{Heq}
  H = -\frac{1}{2}\alpha v^2 +g\,v + h\;,
\end{equation}
where we have introduced ${v}$-independent functions ${g(u,{\bf x})}$ and ${h(u,{\bf x})}$.

In the following we want to determine an analytical expression for $\Sigma$, in order to do that
 we substitute the result \eqref{Heq} into \eqref{r3},
\begin{equation}\label{r6}
2\frac{\Sigma_{,\rho\rho}}{\Sigma}-\frac{(\Sigma_{,\rho})^2}{\Sigma^2}=-\Lambda\Sigma^2+\frac{\vrho}{\Sigma^2}+\alpha\;.
\end{equation}
When we add  the expression \eqref{r4} to \eqref{r6}, we obtain that
\begin{equation}\label{r7}
\Sigma_{,\rho\rho\rho}=-2\Lambda\Sigma^2{\Sigma_{,\rho}}+\alpha{\Sigma_{,\rho}}\;.
\end{equation}
We can rewrite the previous equation as $
\Sigma_{,\rho\rho\rho}=-\frac{2}{3}\Lambda(\Sigma^3)_{,\rho}+\alpha{\Sigma_{,\rho}}$ to be able to integrate it again as
\begin{equation}\label{r8}
\Sigma_{,\rho\rho}=-\frac{2}{3}\Lambda\Sigma^3+\alpha\Sigma+\frac12\beta\;,
\end{equation}
which we can rewrite as
\begin{equation}\label{r9}
\frac12[(\Sigma_{,\rho})^2]_{,\rho}=-\frac{1}{6}\Lambda(\Sigma^4)_{,\rho}+\alpha(\Sigma^2)_{,\rho}+\frac12\beta\Sigma_{,\rho}\;.
\end{equation}
After another integration we get the final formula for the derivative of the function $\Sigma$,
\begin{equation}\label{r10}
(\Sigma_{,\rho})^2=-\frac{1}{3}\Lambda\Sigma^4+\alpha\Sigma^2+\beta\Sigma+c\;,
\end{equation}
and it can be rewritten using \eqref{s3q}  as
\begin{equation}\label{r11}
\gamma S=\left[-\frac{1}{3}\Lambda\Sigma^4+\alpha\Sigma^2+\beta\Sigma+c\right]^{1/2}\;,
\end{equation}
where $\alpha$, $\beta$ and $c$ are integration constants which should be determined.

Furthermore, we are able to determine the constant $c$ explicitly.
When we substitute the result \eqref{r10} and \eqref{r8} into \eqref{r6} we
immediately obtain that $c=-\vrho.$ The constants $\alpha$ and $\beta$ will be determined in the section \ref{ssc:inter}.

\subsection{The Einstein--Maxwell equations for the sources}\label{sources}
The remaining nontrivial components of the Einstein equations are those involving the gyraton source \eqref{mm7}.
To write the source equation we have to evaluate the component $G_{uv}$ using the expressions for derivatives of $\Sigma$.
Then the component $G_{uv}$ has the explicit form
\begin{equation}
G_{uv}=\Lambda\Sigma^2+\frac{\vrho}{\Sigma^2}.
\end{equation}

The $ui$-components give equations related to ${\bf j}$,
\begin{equation}\label{jieqpott}
  \Sigma^2\,{\bf j} = \frac12\rot(\Sigma^4\, b) + \Sigma^2\trgrad g-\alpha\Sigma^2{\bf a}+2\vrho(\vs-{\bf a})\;,
\end{equation}
where $b = \rot{\bf a}\;.$

It is useful to split the source equation into divergence and rotation parts:
\begin{align}
  \trdiv{(\Sigma^2\, {\bf j})}&=\trdiv {\Sigma^2(\trgrad g-\alpha\,{\bf a})},\label{divjeqq}\\
  \rot(\Sigma^2 {\bf j})&= - \frac12 \laplace(\Sigma^4 b) + \rot(\Sigma^2\trgrad g)-\alpha\rot(\Sigma^2{\bf a})-2\vrho\,b\;.\label{rotjeqq}
\end{align}
These are coupled equations for ${g}$ and ${{\bf a}}$. We will return to them below.

The condition \eqref{gyrcon} for the gyraton source gives,
that the sources ${\bf j}$ must be ${v}$-independent and ${j_u}$ has the structure
\begin{equation}\label{jdecomp1}
  j_u = v\,\trdiv(\Sigma^2 {\bf j}) + \iota\;,
\end{equation}
where $\iota(u,{\bf x})$ is $v$--independent function, see \cite{Kadlecova:2010:PHYSR4:} Eq. 2.51.
The gyraton source \eqref{mm7} is therefore  determined by three \mbox{${v}$-independent} functions ${\iota(u,{\bf x})}$ and ${j(u,{\bf x})}$.

The $uu$-component of the Einstein equation gives
\begin{equation}\label{jueq1}
\begin{split}
  j_u =\,&v\left[\trdiv {(\Sigma^2\trgrad g)}-\alpha\trdiv(\Sigma^2 {\bf a})\right]
      +\Sigma^2(\laplace h - (\Sigma^{-2})_{,\rho}h_{,\rho})\\
      &+\frac12\Sigma^4 b^2+ 2\Sigma^2 {\bf a}{\cdot}\trgrad g+(\partial_u+g)\trdiv(\Sigma^2 {\bf a})-\alpha\Sigma^2{\bf a}^2-2\vrho\,(\vs-{\bf a})^2\;.
\end{split}
\end{equation}
Then we can compare the coefficient in front of ${v}$ with \eqref{divjeqq} and we get consistent structure  with \eqref{jdecomp1}. The nontrivial ${v}$-independent part of \eqref{jueq1} gives the equation for the metric function ${h}$,
\begin{equation}\label{heq1}
\begin{split}
 \Sigma^2&(\laplace h -(\Sigma^{-2})_{,\rho}h_{,\rho})=
      \iota \, -\frac12\Sigma^4 b^2- 2\Sigma^2 {\bf a}{\cdot }\trgrad g\\
      & -(\partial_u+g)\trdiv(\Sigma^2 {\bf a})+\alpha\Sigma^2{\bf a}^2+2\vrho\,(\vs-{\bf a})^2\;.
\end{split}
\end{equation}

Now, let us return to solution of equations \eqref{divjeqq} and \eqref{rotjeqq}. The first equation simplifies if
we use gauge condition
\begin{equation}\label{gaugefix}
\trdiv\bigl(\Sigma^2{\bf a}\bigr)=0\;.
\end{equation}
It can be satisfied due to gauge freedom ${v\to v-\chi}$, ${{\bf a}\to{\bf a}-\trgrad\chi}$, cf.\ the discussion in \cite{Kadlecova:2010:PHYSR4:}.
Such a condition implies the existence of a potential ${\tilde\lambda}$, as $\Sigma^2{\bf a} = \rot\tilde\lambda\;.$

The equation \eqref{divjeqq} now reduces to
\begin{equation}\label{divjeqgauged}
    \trdiv(\Sigma^2\trgrad g-\Sigma^2 {\bf j})=0\;.
\end{equation}
It guarantees the existence of a scalar ${\omega}$ such that
\begin{equation}\label{omegadef}
    \trgrad g =  {\bf j} + \Sigma^{-2}\rot\omega\;.
\end{equation}
However, we have to enforce the integrability conditions
\begin{equation}\label{integrability}
    \rot\trgrad g=0\;,
\end{equation}
which turns out to be the equation for ${\omega}$:
\begin{equation}\label{omegaeq}
    \trdiv\bigl(\Sigma^{-2}\trgrad\omega\bigr) = \rot{\bf j}\;.
\end{equation}
We thus obtained the decoupled equations \eqref{omegadef} and \eqref{omegaeq} which determine the metric function ${g}$.

Substituting $\Sigma^2{\bf a} = \rot\tilde\lambda\;$ and \eqref{omegadef} to \eqref{rotjeqq}, and using identity
$b = \rot\bigl(\Sigma^{-2}\rot\tilde\lambda\bigr)\;,$
we get the decoupled equation for ${\tilde\lambda}$:
\begin{equation}\label{lambdaeqq}
\begin{split}
    &\frac12\laplace\Bigl(\Sigma^4\rot\bigl(\Sigma^{-2}\rot\tilde\lambda\bigr)\Bigr)+2\vrho\rot\bigl(\Sigma^{-2}\rot\tilde\lambda\bigr)
       -\alpha\laplace\tilde\lambda=-\laplace\omega\;.
\end{split}
\end{equation}
It is a complicated equation of the forth order. It can be simplified to an ordinary differential equation if we assume
the additional symmetry properties of the fields, e.g., the rotational symmetry around the axis.
The potential ${\tilde\lambda}$ then determines the metric 1-form ${{\bf a}}$ through $\Sigma^2{\bf a} = \rot\tilde\lambda\;$.

After finding ${\bf a}$ one can solve the field equations for ${{\bf s}}$.
The potential equations give immediately that
\begin{equation}\label{phipott}
  \vs = {\rm d}\varphi\;.
\end{equation}
Substituting to the condition $\trdiv(\vs-{\bf a}) = 0\;$ we get the Poisson equation for ${\varphi}$:
\begin{equation}\label{phieq}
    \laplace\varphi = \trdiv{\bf a}\;.
\end{equation}
Finally, the remaining metric function $h$ is determined by the equation \eqref{heq1}.

%%%%%%%%%%%%%%%%%%%%%%%%%%%%%%%%%%%%%%%%%%%%%%%%%%%%%%%%%%%%%%%%%%%%%%%%%%%%%%%%%%%%%%%%%%%%%
\section{The interpretation of the solutions}\label{sc:interprett}
%%%%%%%%%%%%%%%%%%%%%%%%%%%%%%%%%%%%%%%%%%%%%%%%%%%%%%%%%%%%%%%%%%%%%%%%%%%%%%%%%%%%%%%%%%%
\subsection{The geometries of the transversal spacetime}\label{ssc:inter}
In this section we will investigate the geometry of the transversal metric ${\bf q}$ (the wave fronts) \eqref{trmetric4} and
we will determine the constants $\alpha,\,\beta$ in the final equation \eqref{r10}. Subsequently, we
will discuss the various geometries of ${\bf q}$ in proper parametrization and we will determine
the meaning of the parameter $\gamma$.

We impose conditions to the derivatives of $\Sigma$ (i.e., $S$) \eqref{r10}, \eqref{r8} and \eqref{r7} while
using the relation \eqref{s3q} between $\Sigma_{,\rho}$ and $S$ to determine $\alpha$ and $\beta$.

First, we impose conditions at the axis $\rho=0$.
We assume that $S$ and $\Sigma_{,\rho}$ vanish at the axis $\rho=0$, $
S=0,\Sigma_{,\rho}=0,$
second, we can always rescale the metric \eqref{trmetric4} to get $
\Sigma|_{\rho=0}=1,$
third, we want no conical  singularities there, therefore we assume $
\Sigma_{,\rho\rho}|_{\rho=0}=\gamma,$
which we can be justified by computation of the ratio of the circumference $o$ divided by $2\pi$ times radius
in limit $\rho\rightarrow 0$,
\begin{equation}\label{SS3}
\frac{o}{2\pi r}=\frac{2\pi\frac{S}{\Sigma}}{2\pi\int\Sigma\trgrad \rho}=\frac{1}{\Sigma}\left(\frac{S}{\Sigma}\right)_{,\rho}=\frac{1}{\gamma}\frac{\Sigma_{,\rho\rho}\Sigma-(\Sigma_{,\rho})^2}{\Sigma^3}=1.
\end{equation}
Applying the conditions from last paragraph, we obtain
\begin{equation}\begin{aligned}
-&\frac{1}{3}\Lambda+\alpha+\beta-\vrho=0, -\frac{2}{3}\Lambda+\alpha+\frac12\beta=\gamma.
\end{aligned}\end{equation}

We can then determine the constants $\alpha$ and $\beta$ explicitly
in terms of the cosmological constant $\Lambda$, the density of electromagnetic
field $\vrho$ and the parameter $\gamma$,
\begin{equation}\begin{aligned}\label{glab}
\alpha&=\Lambda-\vrho+2\gamma, \beta=-\frac{2}{3}\Lambda+2\vrho-2\gamma.
\end{aligned}\end{equation}
We can conveniently rewrite \eqref{s3q},
\begin{equation}\begin{aligned}\label{s33}
&(\gamma S)^2=(\Sigma_{,\rho})^2=
=\left[-\frac{1}{3}\frac{\Lambda}{\gamma^2}(\Sigma^2-2)\Sigma-\frac{\vrho}{\gamma^2}(\Sigma-1)+\frac{2}{\gamma}\Sigma\right](\Sigma-1).
\end{aligned}\end{equation}

Now we know explicitly the constants in the derivative of $\Sigma$ and we can
investigate the interpretation of the generalized Melvin spacetime.
It is convenient to introduce  new coordinate $x$ as
\begin{equation}\label{x}
\Sigma=1+\gamma x,
\end{equation}
then we can write that
\begin{equation}\label{sx}
S=x_{,\rho},\quad\Sigma_{,\rho}=\gamma x_{,\rho}\;.
\end{equation}

The transversal metric ${\bf q}$ \eqref{trmetric4} then
can be rewritten as
\begin{equation}\label{q1}
{\bf q}=\left(\frac{\Sigma}{S}\right)^2\trgrad x^2+\left(\frac{S}{\Sigma}\right)^2\trgrad \phi^2=\frac{1}{G}\trgrad x^2+G\trgrad \phi^2,
\end{equation}
where we can express the new function $G$ as
\begin{equation}\label{q2}
G=\left(\frac{S}{\Sigma}\right)^2=-\frac{1}{3}\frac{\Lambda}{\gamma^2}\Sigma^2+\frac{\alpha}{\gamma^2}+\frac{\beta}{\gamma^2}\frac{1}{\Sigma}-\frac{\vrho}{\gamma^2}\frac{1}{\Sigma^2},
\end{equation}
and
\begin{equation}\label{q3}
S^2=\mp \ell^2\gamma^2 x^4 \mp \ell^2\gamma x^3 +(\mp 3\ell^2-\vrho+2\gamma)x^2+2x,
\end{equation}
where we denoted $
\mp \ell^2=\frac{\Lambda}{3}$ where $\pm=\text{sign}\,\Lambda.$

Before we will discuss the possible geometries given by the transversal metric ${\bf q}$
\eqref{trmetric4} and interpret them accordingly
we introduce important characteristics for the generalized Melvin spacetime.

The radial radius is then defined as
\begin{equation}\label{q5}
r=\int_{0}^{x}\frac{1}{\sqrt{G}}\trgrad x,
\end{equation}
the circumference radius is simply given
by the function $G$, $R=\sqrt{G}.$
Interestingly, the ratio of the radia is then determined
by the derivative of $G$,
\begin{equation}\label{q7}
\frac{\trgrad R}{\trgrad r}=\sqrt{G}\frac{\trgrad \sqrt{G}}{\trgrad x}=\frac12 G_{,x}.
\end{equation}

The scalar curvature of ${\bf q}$ can be also written as
\begin{equation}\begin{aligned}\label{RR}
\mathcal{R}=-G_{,xx}&=-\frac{2}{\Sigma^4}\left[3\Sigma_{,\rho}+\frac{2}{3}\Lambda\Sigma^4-3\alpha\Sigma^2-2\beta\Sigma\right].
\end{aligned}\end{equation}

The geometries of the transversal spacetime ${\bf q}$ can be illustrated
by investigating the function $G$ and its roots when we consider different
values of $\Lambda$,\, $\vrho$ and of the parameter $\gamma$.

\begin{figure}[htp]
\begin{center}
\includegraphics{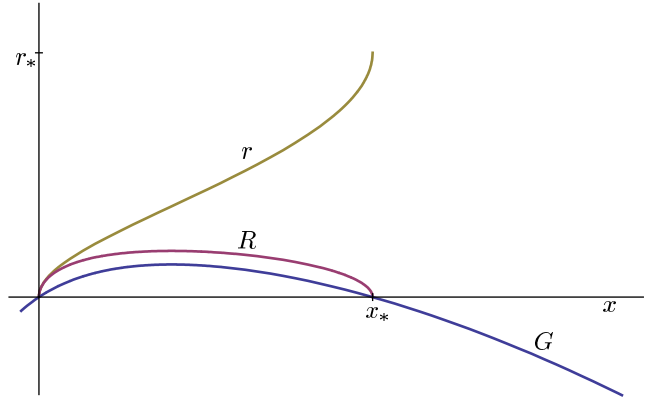}
\end{center}
\caption{\label{fig:graf1}%
The case when $\Lambda>0$ which represents closed spacetime. The function $G$ is visualized for
any value of $\vrho$ and $\gamma$. The coordinate $\rho$ ranges $\rho\in(0,\rho_{*})$ where the $\rho_{*}$ is the first root of $G$ where the spacetime closes.}
\end{figure}

First, we consider positive cosmological constant $\Lambda>0$ for any $\vrho$ and $\gamma$ we obtain {\it closed space} where $\rho\in(0,\rho_{*})$ and $\rho_{*}$ represents the first positive root of $G$ where in fact the spacetime closes itself. The other characteristics are:
 the radial radius tends to a finite value $r\rightarrow r_{*}$ at the $\rho_{*}$  and the circumference radius vanishes $R\rightarrow 0$ when $\rho\rightarrow \rho_{*}$ .
This special case is visualized in the graph \ref{fig:graf1}.

For the vanishing cosmological constant $\Lambda=0$ we obtain three possible
spacetimes according to the values of $\vrho$ and $\gamma$.

\begin{figure}[htp]
\begin{center}
\includegraphics{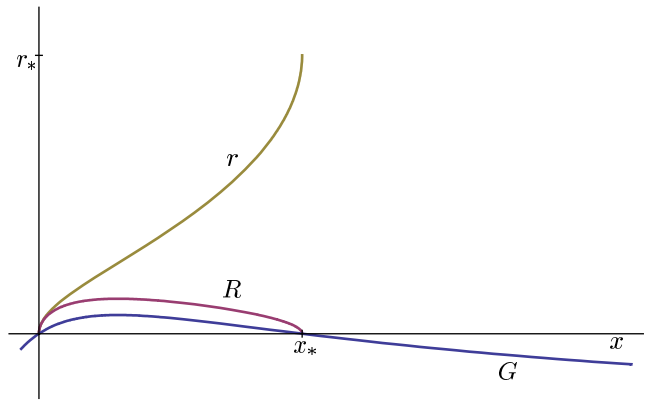}
\end{center}
\caption{\label{fig:graf2}%
The case when $\Lambda=0$ and $\vrho>2\gamma$ represents the closed spacetime. The function $G$ is visualized for $\vrho>2\gamma$ and the coordinate $\rho$ ranges $\rho\in(0,\rho_{*})$ where the $\rho_{*}$ is the root of $G$ where the spacetime closes.}
\end{figure}

When $\vrho>2\gamma$ then we get {\it closed space} where the range of the coordinate $\rho$ goes again
as $\rho\in(0,\rho_{*})$ and $\rho_{*}$ is then the root of $G$ and it is the closing point of the universe.
The radia are then $r\rightarrow r_{*}$ and $R\rightarrow 0$ when $\rho\rightarrow \rho_{*}$, see the graph \ref{fig:graf2}.

\begin{figure}[htp]
\begin{center}
\includegraphics{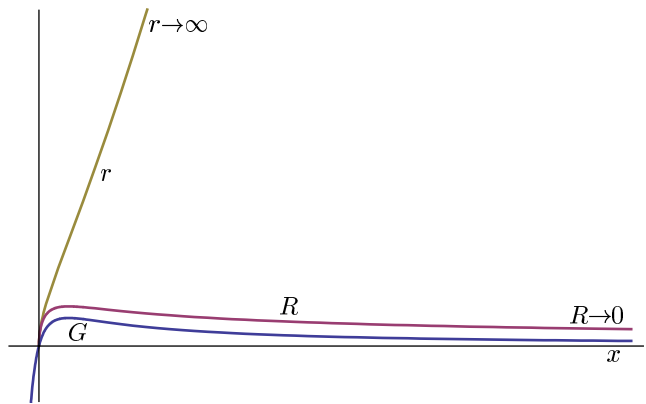}
\end{center}
\caption{\label{fig:graf3}%
The case when $\Lambda=0$ and $\vrho=2\gamma$ then  represents the closed spacetime with an infinite peak. The function $G$ is visualized for $\vrho=2\gamma$ and the coordinate $\rho$ ranges $\rho\rightarrow\infty$.}
\end{figure}

When $\vrho=2\gamma$ then we obtain {\it closed space with and infinite peak} for $\rho\rightarrow \infty$.
Therefore, when $\rho\rightarrow \infty$ the radial radius tends to infinity $r\rightarrow \infty$ and the circumference radius goes to zero $R\rightarrow 0$, see the graph \ref{fig:graf3}.
This case represents the pure Melvin spacetime \cite{Bonnor:1954:PRS:,Melvin:1965:PHYSR:} which we discussed in \cite{Kadlecova:2010:PHYSR4:}.

\begin{figure}[htp]
\begin{center}
\includegraphics{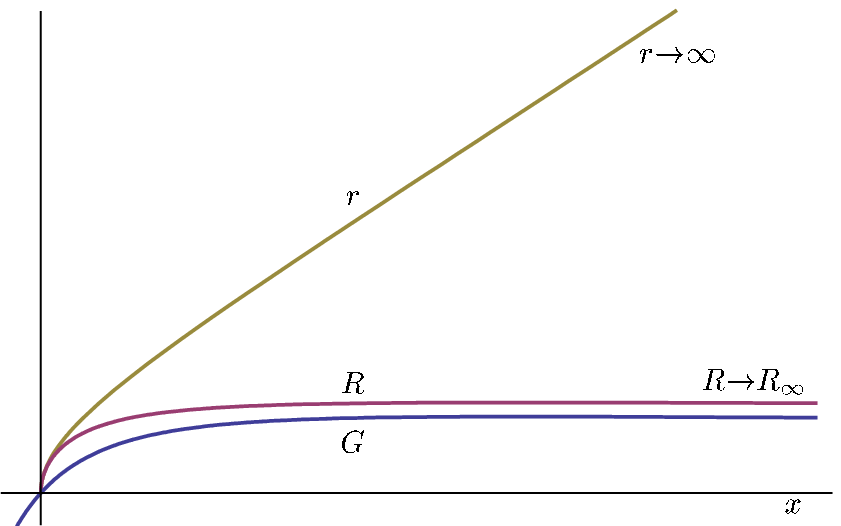}
\end{center}
\caption{\label{fig:graf4}%
The case when $\Lambda=0$ and $\vrho<2\gamma$ then  represents the open spacetime. The function $G$ is visualized for $\vrho<2\gamma$ and the coordinate $\rho$ ranges $\rho\rightarrow\infty$.}
\end{figure}

When $\vrho<2\gamma$ then we obtain {\it an open space} for $\rho\in(0,\infty)$.
When $\rho\rightarrow \infty$, the radial radius tends to infinity $r\rightarrow \infty$; however, the circumference radius goes to a finite value, $R\rightarrow R_{\infty}$, see the graph \ref{fig:graf4}.

\begin{figure}[htp]
\begin{center}
\includegraphics{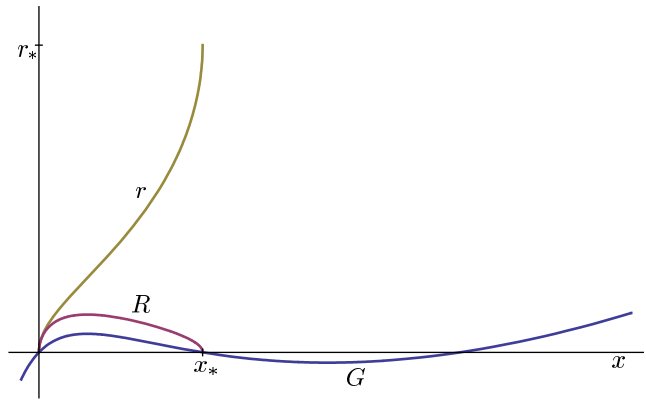}
\end{center}
\caption{\label{fig:graf5}%
The case when $\Lambda<0$ and $\gamma<\gamma_{\mathrm cr}$ represents the closed spacetime. The coordinate $\rho$ ranges $\rho\in(0,\rho_{*})$ where the $\rho_{*}$ is the root of $G$ where the spacetime closes.}
\end{figure}

When we consider the negative cosmological constant $\Lambda<0$ we obtain three possible
spacetimes according to the values of $\gamma$.
For ${\gamma}$ smaller than certain critical value ${\gamma_{\mathrm cr}}$ (which depends on ${\Lambda}$ and ${\vrho}$), we get {\it closed space} where the range of the coordinate $\rho$ goes again as $\rho\in(0,\rho_{*})$ and $\rho_{*}$ is then the root of $G$ and the closing point of the universe.
The radia are then $r\rightarrow r_{*}$ and $R\rightarrow 0$ when $\rho\rightarrow \rho_{*}$, see the graph \ref{fig:graf5}.

\begin{figure}[htp]
\begin{center}
\includegraphics{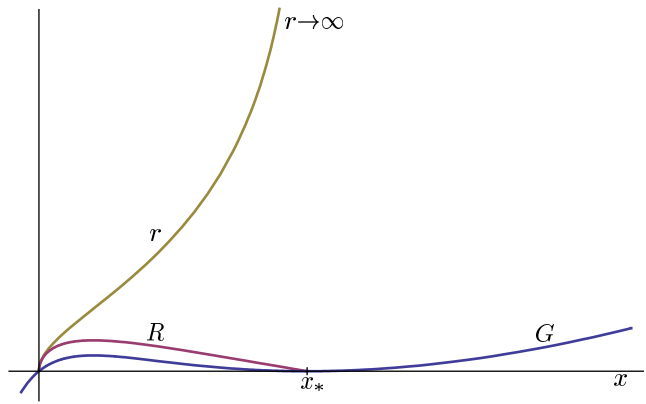}
\end{center}
\caption{\label{fig:graf6}%
The case when $\Lambda<0$ and $\gamma=\gamma_{\mathrm cr}$ represents the asymptotically closed spacetime. The coordinate $\rho$ ranges $\rho\in(0,\rho_{*})$ where the $\rho_{*}$ is the root of $G$. The radial distance tends to infinity and the circumference shrinks to zero.}
\end{figure}

When $\gamma=\gamma_{\mathrm cr}$, we obtain {\it closed space with and infinite peak} where the range of the coordinate $\rho$ goes
as $\rho\in(0,\rho_{*})$ and $\rho_{*}$ is the root of $G$.
The radia are then $r\rightarrow\infty$ and $R\rightarrow 0$ when $\rho\rightarrow \rho_{*}$, see the graph \ref{fig:graf6}.

\begin{figure}[htp]
\begin{center}
\includegraphics{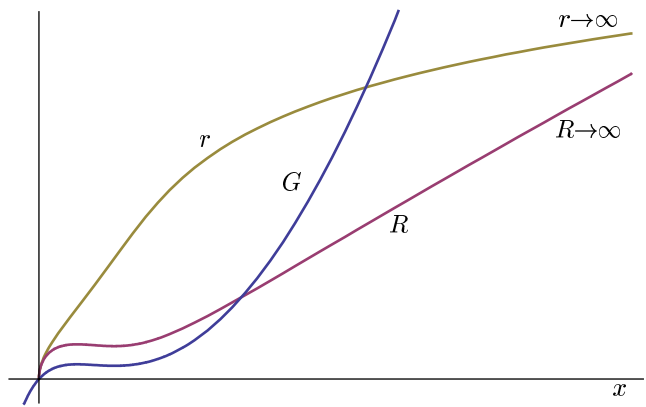}
\end{center}
\caption{\label{fig:graf7}%
The case when $\Lambda<0$ and $\gamma>\gamma_{\mathrm cr}$ represents the open spacetime. The coordinate ${\rho}$ takes positive real values.
For $\rho\rightarrow \infty$,\,$r\rightarrow \infty$, and $R\rightarrow R_{\infty}$, see the graph \ref{fig:graf7}.}
\end{figure}

When $\gamma>\gamma_{\mathrm cr}$, we obtain {\it open space} for $\rho\in(0,\infty)$.
For $\rho\rightarrow \infty$,\,$r\rightarrow \infty$, and $R\rightarrow R_{\infty}$, see the graph \ref{fig:graf7}.

\begin{table}
\caption{\label{table16} Possible geometries of the transversal spacetime ${\bf q}$. Here $\Lambda$ is a cosmological constant, $\vrho$ is energy density of the electromagnetic field and $\gamma$ is the parameter of `Melviniztion' of the spacetime. Critical value ${\gamma_{\mathrm cr}(\Lambda,\vrho)}$} is determined by the condition that the function ${G}$ has degenerated root at ${\rho_*}$.
%\begin{ruledtabular}
\begin{tabular}{cccccc}
 $\Lambda$ &  $\vrho$,$\gamma$ & \text{transversal spacetime}  & $\rho$ & $r|_{\rho\rightarrow\rho_{*}}$ & $R|_{\rho\rightarrow\rho_{*}}$ \\
\hline
 $\Lambda>0$ & \text{any} & closed space       & $(0,\rho_{*})$ & $r_{*}$& $0$ \\
 \hline
 & $\gamma<\vrho/2$   & closed space         & $(0,\rho_{*})$ & $r_{*}$& $0$ \\
 $\Lambda=0$ & $\gamma=\vrho/2$    & Melvin universe  & $\mathbb{R}^{+}$ & ${\infty}$ & $0$ \\
 & $\gamma>\vrho/2$   & open space  & $\mathbb{R}^{+}$ & $\infty$ & $R_{\infty}$\\
 \hline
 & $\gamma<\gamma_{\mathrm cr}$   & closed space  & $(0,\rho_{*})$ & $r_{*}$ & $0$  \\
$\Lambda<0$ &  $\gamma=\gamma_{\mathrm cr}$    & closed with $\infty$ peak & $(0,\rho_{*})$ & $\infty$ & $0$ \\
  &  $\gamma>\gamma_{\mathrm cr}$    & open space & $\mathbb{R}^{+}$ & $\infty$ & $\infty$\\
\end{tabular}
%\end{ruledtabular}
\end{table}

We have summarized our resulting geometries arising from the generalized Melvin universe in a Table \ref{table16}.

To conclude this section, we have investigated the transversal spacetime of the generalized Melvin universe.
We have identified the constants $\alpha$ and $\beta$, interpreted them in terms of the cosmological constant $\Lambda$, $\vrho$ and $\gamma$.
After suitable parametrization of the transversal spacetime we have discussed all possible cases of universes which are contained
in the generalized Melvin universe. The Melvin universe occurs as a special case. We have visualized these cases in graphs and summarized them in the Table \ref{table16}.

The parameter $\gamma$ changes the character of the  influence of the electromagnetic field on the geometry. With larger ${\gamma}$ the influence is stronger and for $\Lambda \leq 0$ it can even change the global structure of the spacetime, what exactly happens for the critical value ${\gamma_{\mathrm cr}}$ (for $\Lambda=0$ $\gamma_{\mathrm cr}=\vrho/2$).

%%%%%%%%%%%%%%%%%%%%%%%%%%%%%%%%%%%%%%%%%%%%%%%%%%%%%%%%%%%%%%%%%%%%%%
\subsection{The backgrounds for our solutions}
The background spacetimes are defined as a limit when $h=g=0$ and ${\bf a}=0$, then
the metric \eqref{s1w} reduces to
\begin{equation}\label{ss1}
{\bf g}={\bf q}-\Sigma^2\trgrad u\vee\trgrad v +\alpha v^2\Sigma^2\trgrad u^2.
\end{equation}

The metric \eqref{ss1} admits one killing vector $\partial_{\phi}$
which corresponds to cylindrical symmetry.

Using the adapted null tetrad ${\bf k}=\partial_{v},\,{\bf l}=\Sigma^{-2}(\partial_{u}+\frac12\alpha v^2\partial_{v}),\,{\bf m}=\frac{1}{\sqrt{2}}(\Sigma^{-1}\partial_{\rho}-i\Sigma S^{-1}\partial_{\phi})$, the only non--vanishing components of Weyl and Ricci tensors are,
\begin{equation}\begin{aligned}\label{Melpsii}
\Psi_{2}&=\frac{1}{2\Sigma^4}\left(\beta\Sigma-2\vrho\right), 
\Phi_{11}&=\frac{1}{2\Sigma^4}\vrho.
\end{aligned}\end{equation}
This demonstrates that the generalized Melvin universe is a non--vacuum solution
of type D, except the points where $\Psi_{2}=0$.

The background metric \eqref{ss1} contains several sub--solutions.
For $\Lambda=0$ and $\vrho=2\gamma$ we obtain the Melvin universe which
serves as a background in \cite{Kadlecova:2010:PHYSR4:} and the
the only non--vanishing Weyl and curvature scalars are
\begin{equation}\begin{aligned}\label{Mel1}
\Phi_{2}&=-\frac{\vrho}{2\Sigma^4}(2-\Sigma)=\frac{1}{2}\frac{\vrho}{\Sigma^4}(-1+\frac{1}{4}\vrho\rho^2),
\Psi_{11}&=\frac{1}{2\Sigma^4}\vrho,
\end{aligned}\end{equation}
where we have used the $\Sigma=1+\frac{1}{4}\vrho\rho^2$ which specifies the Melvin
spacetime.
The scalar curvature of the transversal spacetime ${\bf q}$ \eqref{RR} then
becomes
\begin{equation}
\mathcal{R}=0,
\end{equation}
which agrees with \cite{Kadlecova:2010:PHYSR4:}.

For $\Sigma=1$ we get the direct product background
spacetimes, the metric \eqref{ss1} reduces to
\begin{equation}\label{ss2}
{\bf g}={\bf q}-\trgrad u\vee\trgrad v +\alpha v^2\trgrad u^2,
\end{equation}
the only non--vanishing Weyl and curvature scalars then are
\begin{equation}\begin{aligned}\label{Mel2}
\Psi_{2}=\frac{1}{2}\left(\beta-2\vrho\right)=-\frac{\Lambda}{3},\,\Phi_{11}=\frac{1}{2}\vrho.
\end{aligned}\end{equation}
The scalar curvature of the transversal spacetime ${\bf q}$ \eqref{RR} then
becomes
\begin{equation}
\mathcal{R}=2(\Lambda+\vrho),
\end{equation}
which agrees with \cite{Kadlecova:2009:PHYSR4:}.

\begin{table}
\caption{\label{table2}Some of possible background spacetimes in the case $\gamma=0$ which  represents the direct product of two 2-spaces of constant curvature. The parameter
$\Lambda_{+}=\Lambda+\vrho$ gives the geometry of the wave front and $\Lambda_{-}=\Lambda-\vrho$  determines the conformal structure of the background.}
%\begin{ruledtabular}
\begin{tabular}{cccccccc}
 $\Lambda_{+}$ &  $\Lambda_{-}$ & \text{geometry} & \text{background}  & $\Lambda$ & $\vrho$\\
\hline
 0 & 0                  & ${E^{2}\times M_{2}}$  & Minkowski        & $=0$ & $=0$ \\
 $\Lambda$ & $\Lambda$  & ${S^{2}\times dS_{2}}$  & Nariai          & $>0$ & $=0$ \\
 $\Lambda$ & $\Lambda$  & ${H^{2}\times AdS_{2}}$  & anti-Nariai   & $<0$ & $=0$ \\
 $\vrho$ & $-\vrho$ & ${S^{2}\times AdS_{2}}$  & Bertotti--Robinson   & $=0$ & $>0$ \\
 $2\Lambda$ & 0   & ${S^{2}\times M_{2}}$  & Pleba\'{n}ski--Hacyan   & $>0$ & $=\Lambda$ \\
 0 &  $2\Lambda$  & ${E^{2}\times AdS_{2}}$  & Pleba\'{n}ski--Hacyan & $<0$ & $=|\Lambda|$
\end{tabular}
%\end{ruledtabular}
\end{table}

To summarize the background metric \eqref{ss1} generalizes the metric for the pure  Melvin universe and
the direct product spacetimes into one background metric and combines their
properties.
%%%%%%%%%%%%%%%%%%%%%%%%%%%%%%%%%%%%%%%%%%%%%%%%%%%%%%%%%%%%%%%%%%%%%%%%%%%%%%%%%%
\section{The scalar polynomial invariants}\label{sc:Invariants}
%%%%%%%%%%%%%%%%%%%%%%%%%%%%%%%%%%%%%%%%%%%%%%%%%%%%%%%%%%%%%%%%%%%%%%%%%%%%%%%%%%
The scalar invariants are important characteristics of gyraton spacetimes.
The gyratons in the Minkowski spacetime \cite{Fro-Is-Zel:2005:PHYSR4:} have vanishing invariants (VSI) \cite{Prav-Prav:2002:CLAQG:}, the gyratons in the AdS \cite{Fro-Zel:2005:PHYSR4:} and direct
product spacetimes \cite{Kadlecova:2009:PHYSR4:} have all invariants constant (CSI) \cite{Coley-Her-Pel:2006:CLAQG}.
The invariants are independent of all metric functions $a_{i}$ which characterize the gyraton, and have the same values as the corresponding invariants of the background spacetime.
We have shown that similar property is valid also for the gyraton on Melvin spacetime \cite{Kadlecova:2010:PHYSR4:}, but the invariants
are functions of the coordinate $\rho$ and depend on the constant density $\vrho$.

In these cases, the invariants are independent of all metric functions
which characterize the gyraton, and have the same
values as the corresponding invariants of the background spacetime.
We observed that similar property is valid also for the gyraton on
Melvin spacetime and it is valid also for its generalization with $\Lambda$, however, in this case the invariants are generally \emph{non-constant},
namely, they depend on the coordinate $\rho$. This property is a consequence of general theorem holding for the relevant subclass of Kundt solution, see Theorem II.7 in \cite{ColeyEtal:2010}. For more details, see \cite{Kadlecova:2013:}.

%%%%%%%%%%%%%%%%%%%%%%%%%%%%%%%%%%%%%%%%%%%%%%%%%%%%%%%%%%%%%%%%%%%%%%%%%%%%%%%%%%%%%%%%%%%%%
\section{Conclusion}\label{sc:conclusionE}
%%%%%%%%%%%%%%%%%%%%%%%%%%%%%%%%%%%%%%%%%%%%%%%%%%%%%%%%%%%%%%%%%%%%%%%%%%%%%%%%%%%%%%%%%%%%%
Our work generalizes the studies of the gyraton on the Melvin universe \cite{Kadlecova:2010:PHYSR4:}.
Namely we have generalized the transversal background metric for the pure Melvin universe where instead of the coordinate $\rho$ we have assumed general function $S$ dependent only on the coordinate $\rho$. This change enabled us to find new solutions with possible non--zero cosmological constant. This is not allowed for the pure Melvin background spacetime.
We were able to derive relation between metric functions $\Sigma$ and $S$ from the source free  Einstein--Maxwell equations.
The derivative of the function $\Sigma_{,\rho}$ is then polynomial in the function $\Sigma$ itself and contains four parameters. We have showed that these parameters can be expressed using constants $\Lambda$, $\vrho$ and $\gamma$.

The Einstein--Maxwell equations reduce again to the set of linear equations on the 2--dimensional transverse spacetime which has non--trivial geometry given by the generalized Melvin spacetime \eqref{trmetric4}. Fortunately, these equations do decouple and they can be solved least in principle for any distribution of the matter sources.

In detail, we have studied the transversal geometries of  generalized Melvin spacetime \eqref{trmetric4}.
We have discussed the various possible values of constants $\Lambda$, $\vrho$ and $\gamma$. It occurs that for $\Lambda>0$ the transversal geometry represents only one type of space, the case $\Lambda=0$ includes three different spaces, one of them corresponds to the Melvin spacetime as a special case. The case $\Lambda<0$ also describes three types of spaces. We have visualized them in several graphs in Section \ref{sc:interprett}
and summarized them in the Table \ref{table16}.
Thanks to this discussion we were able to interpret the parameter $\gamma$ as the parameter which makes the electromagnetic field of the direct product spacetimes stronger.

We have investigated the polynomial scalar invariants. In this generalized case, the invariants are again
not constant and they are functions of the metric function $\Sigma$ and the full gyratonic
metric has the same invariants as the background metric.

% Acknowledgements are created using the command \ack:
\ack%%%%%%%%%%%%%%%%%%%%%%%%%%%%%%%%%%%%%%%%%%%%%%%%%%%%%%%%%%%%%%%%%%%%%%%

% Contributors involved in `Vyzkumny zamer' can use macro \InstResCode
% instead of specifying the alphanumerical code explicitly:
The present work was supported by the grant GAUK 12209 by the Czech Ministry of
Education, the project SVV 261301 of the Charles University in Prague and
the LC06014 project of the Center of Theoretical Astrophysics.

% Here we specify the basename of the bibliography database file,
% in this case \jobname=ragsamp:
\bibliography{mynames}

\begin{thebibliography}{11}
\expandafter\ifx\csname natexlab\endcsname\relax\def\natexlab#1{#1}\fi
\expandafter\ifx\csname url\endcsname\relax
  \def\url#1{\texttt{#1}}\fi
\expandafter\ifx\csname urlprefix\endcsname\relax\def\urlprefix{URL }\fi
\providecommand{\selectlanguage}[1]{\relax}
\providecommand{\eprint}[2][]{\url{#2}}

\bibitem[{Bonnor(1954)}]{Bonnor:1954:PRS:}
Bonnor, W.~B. (1954), {The Equations of Motion in the Non-Symmetric Unified
  Field Theory}, \emph{Proc. Roy. Soc. London A}, \textbf{67}(225).

\bibitem[{Coley et~al.(2006)Coley, Hervik and
  Pelavas}]{Coley-Her-Pel:2006:CLAQG}
Coley, A.~A., Hervik, S. and Pelavas, N. (2006), {On spacetimes with constant
  scalar invariants}, \emph{Class. Quant. Gravity}, \textbf{23}, pp.
  3053--3074.

\bibitem[{Coley et~al.(2010)Coley, Hervik and Pelavas}]{ColeyEtal:2010}
Coley, A.~A., Hervik, S. and Pelavas, N. (2010), {Lorentzian manifolds and
  scalar curvature invariants}, \emph{Class. Quant. Gravity}.

\bibitem[{Frolov et~al.(2005)Frolov, Israel and
  Zelnikov}]{Fro-Is-Zel:2005:PHYSR4:}
Frolov, V.~P., Israel, W. and Zelnikov, A. (2005), {Gravitational field of
  relativistic gyratons}, \emph{Phys. Rev. D}, \textbf{72}, p. 084031.

\bibitem[{Frolov and Zelnikov(2005)}]{Fro-Zel:2005:PHYSR4:}
Frolov, V.~P. and Zelnikov, A. (2005), {Relativistic gyratons in asymptotically
  AdS spacetime}, \emph{Phys. Rev. D}, \textbf{72}, p. 104005.

\bibitem[{Kadlecov\'{a}(2013)}]{Kadlecova:2013:}
Kadlecov\'{a}, H. (2013), {Gravitational field of gyratons on various
  background spacetimes, arXiv:1308.5008}.

\bibitem[{Kadlecov\'{a} and Krtou\v{s}(2010)}]{Kadlecova:2010:PHYSR4:}
Kadlecov\'{a}, H. and Krtou\v{s}, P. (2010), {Gyratons on Melvin spacetime},
  \emph{Phys. Rev. D}, \textbf{82}, p. 044041.

\bibitem[{Kadlecov\'{a} et~al.(2009)Kadlecov\'{a}, Zelnikov, Krtou\v{s} and
  Podolsk\'{y}}]{Kadlecova:2009:PHYSR4:}
Kadlecov\'{a}, H., Zelnikov, A., Krtou\v{s}, P. and Podolsk\'{y}, J. (2009),
  {Gyratons on direct--product spacetimes}, \emph{Phys. Rev. D}, \textbf{80},
  p. 024004.

\bibitem[{Melvin(1965)}]{Melvin:1965:PHYSR:}
Melvin, M.~A. (1965), {Dynamics of Cylindrical Electromagnetic Universes},
  \emph{Phys. Rev.}, \textbf{139}, pp. B225--B243.

\bibitem[{Pravda et~al.(2002)Pravda, Pravdova, Coley and
  Milson}]{Prav-Prav:2002:CLAQG:}
Pravda, V., Pravdova, A., Coley, A. and Milson, R. (2002), {All spacetimes with
  vanishing curvature invariants}, \emph{Class. Quant. Gravity}, \textbf{19},
  pp. 6213--6236.

\bibitem[{Stephani et~al.(2003)Stephani, Kramer, Maccallum, Hoenselaers and
  Herlt}]{Step:2003:Cam:}
Stephani, H., Kramer, D., Maccallum, M., Hoenselaers and Herlt, E. (2003),
  \emph{{Exact Solutions of Einstein's Field Equations}}, Cambridge University
  Press, Cambridge.

\end{thebibliography}
\end{document}